\documentclass[copyright]{eptcs}
\providecommand{\event}{QAPL2011} 

\usepackage{graphicx}
\usepackage{multirow}
\usepackage{booktabs}

\title{QuantUM: Quantitative Safety Analysis of UML Models}
\author{Florian Leitner-Fischer \qquad\qquad Stefan Leue
\institute{University of Konstanz,
Germany}
\email{\quad florian.leitner@uni-konstanz.de \quad\qquad stefan.leue@uni-konstanz.de}
}
\def\titlerunning{QuantUM: Quantitative Analysis of UML Models}
\def\authorrunning{F. Leitner-Fischer \& S. Leue}
\begin{document}
\maketitle

\begin{abstract}
When developing a safety-critical system it is essential to obtain an assessment of different design alternatives. In particular, an early safety assessment of the architectural design of a system is desirable. In spite of the plethora of available formal quantitative analysis methods it is still difficult for software and system architects to integrate these techniques into their every day work. This is mainly due to the lack of methods that can be directly applied to architecture level models, for instance given as UML diagrams. Also, it is necessary that the description methods used do not require a  profound knowledge of formal methods. Our approach bridges this gap and improves the integration of quantitative safety analysis methods into the development process. All inputs of the analysis are specified at  the level of a UML model. This model is then automatically translated into the analysis model, and the results of the analysis are consequently represented on the level of the UML model. Thus the analysis model and the formal methods used during the analysis are hidden from the user. We illustrate the usefulness of our approach using an industrial strength case study.
\end{abstract}

\section{Introduction}\label{sec:Introduction}
In a recent joint work with our industrial partner TRW Automotive GmbH  
we have proven the applicability of probabilistic verification techniques to safety analysis
in an industrial setting~\cite{AljazzarL09a}.  The analysis approach that we used was that of 
probabilistic Failure Modes Effect Analysis (pFMEA)  \cite{Grunske07}.

The most notable shortcoming of the approach that we observed lies 
in the missing connection of our analysis to existing high-level architecture models
and the modeling languages that they are typically written in.  
TRW Automotive GmbH, like many other software development enterprises in the embedded systems domain, 
mostly uses the Unified Modeling Language (UML) \cite{umlspec}  
for system modeling.
During the pFMEA we however had to use the language provided by the analysis tool that we used,
in this case the input language of the stochastic model checker PRISM~\cite{HintonKNP06}. 
This required a manual translation from the design language UML to the formal modeling language PRISM.
This manual translation has the following shortcomings: 
(1) It is a time-consuming and hence expensive process. 
(2) It is error-prone, since behaviors may be introduced that are not present in the original model. 
(3) The results of the formal analysis may not be easily transferable to the level of the high-level 
design language. To avoid problems that may result from (2) and (3), 
additional checks for plausibility have to be made, which again consume time.
Some introduced errors may even remain unnoticed. 

The objective of this paper is to bridge the gap between architectural design and 
formal stochastic modeling languages so as to remedy the negative implications of this gap listed above. 
This allows for a more seamless integration of formal dependability and reliability analysis into
the design process. We propose an extension of the UML to capture probabilistic and error 
behavior information that are relevant for a formal stochastic analysis, such as when performing,
for instance, pFMEA. All inputs of the analysis can be specified at the level of the UML model. 
In order to achieve this goal, we present an extension of the UML that allows for
the annotation of UML models with quantitative information, such as for instance failure rates. 
Additionally, a translation process from UML models to the PRISM language is defined.

\paragraph{Structure of the paper.}
The remainder of the paper is structured as follows:
In Section \ref{sec:QuantUM} we present our QuantUM approach,
in  Section \ref{sec:CSL} we discuss the automatic construction of CSL formulas. The mapping
of probabilistic counterexamples onto UML sequence diagrams is described in Section \ref{sec:Sequence}.
Section \ref{sec:CaseStudy} is devoted to the case study of an airbag system. 
Followed by a discussion of related work in Section \ref{sec:RelatedWork}. 
We conclude in Section \ref{sec:Conclusion}.

\section{The QuantUM Approach}\label{sec:QuantUM}
In our approach all inputs of the analysis are specified at the level of a UML model. 
To facilitate the specification, we propose a quantitative extension of the UML.
This extension is defined in terms of an UML Profile and takes advantage of the
UML concept of a stereotype. 
Due to space restrictions, we can not elaborate all details of the proposed profile here,
we will only present two central concepts, namely the \emph{QUMComponent} stereotype 
and how the operational profile is captured with UML state machine diagrams.
A comprehensive description of the profile can be found in \cite{mscflf}.

The stereotype \emph{QUMComponent} can be assigned to all UML elements
that represent building blocks of the real system, which include classes, components and interfaces.
Each element with the stereotype \emph{QUMComponent} comprises 
up to one (hierarchical) state machine representing the normal behavior and between 
one and finitely many (hierarchical) state machines representing possible failure patterns. 
These state machines can be either state machines that are especially constructed for 
this  \emph{QUMComponent}, or they can be taken from a repository of state machines describing standard failure behaviors.
The repository provides state machines for all standard components (e.g., switches) and the reuse of these 
state machines saves modeling effort and avoids redundant descriptions. 
In some cases, the normal behavior of a \emph{QUMComponent} is not of interest for the analysis, for instance when describing 
failures of  external components. In those cases the specification of the failure pattern state machines is sufficient.
In addition, each \emph{QUMComponent}  
comprises a list called \emph{Rates} that contains rates together with names identifying them. 

In order to capture the operational profile and particularly to allow the specification of quantitative information,
such as failure rates, we extend the \emph{Transition} element used in UML state machines with the stereotypes 
\emph{QUMAbstractStochasticTransition} and \emph{QUMStochasticTransition}. 
These stereotypes allow the user to specify transition rates as well as a name for the transition.
The specified rates are used as transition rates for the continuous-time Markov chains that are 
generated for the purpose of stochastic analysis.
Transitions with the stereotype \emph{QUMAbstractStochasticTransition} are transitions that do not have a default rate. 
If a state machine is connected to a  \emph{QUMComponent} element,
there has to be a rate in the \emph{Rates} list of the \emph{QUMComponent}  that has the same name as the \emph{QUMAbstractStochasticTransition}. 
This rate is then considered for the \emph{QUMAbstractStochasticTransition}. 
The \emph{QUMAbstractStochasticTransition} allows to define general state machines in a repository
where the rates can be set individually for each component.  
The stereotypes \emph{QUMAbstractFailureTransition}, \emph{QUMAbstractRepairTransition}, 
 \emph{QUMFailureTransition} and \emph{QUMRepairTransition} are specializations of
\emph{QUMAbstractStochasticTransition} and \emph{QUMStochasticTransition}, respectively.

\begin{figure}[htb]
\centering
\includegraphics[width=12cm]{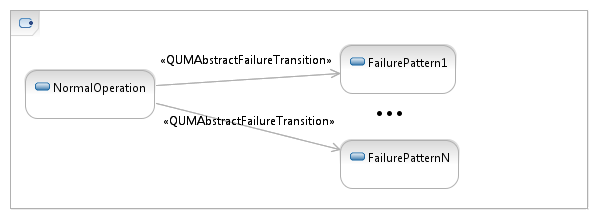}
\caption{Combination of normal behavior and failure pattern state machines.}
\label{fig:QuantUM_Example_StateMachine}
\end{figure}

The normal behavior state machine and all failure pattern state machines of a \emph{QUMComponent} are implicitly combined in one 
hierarchical state machine, cf.\ Figure \ref{fig:QuantUM_Example_StateMachine}. The combined state machine 
is automatically generated by the analysis tool and is not visible to the user. 
Its semantics can be described as follows: initially, the component executes the normal behavior state machine. 
If a \emph{QUMAbstractFailureTransition} is enabled, the component 
will enter the state machine describing the corresponding failure pattern
with the specified rate.
The decision, which of the $n$ FailurePatterns is selected, is made by a stochastic "race"~between the transitions.

We define the semantics of QuantUM by defining a set of translation rules which map  
the UML artifacts that we defined to the input language of the 
model checker PRISM \cite{HintonKNP06}, which is considered to possess a formal semantics. 
This corresponds to the commonly held idea that the 
semantics of a UML model is largely defined by the underlying code generator. 
These translation rules enable a fully automated translation of the UML model into the PRISM
language.
We base our semantic transformation on the operational UML semantics defined in \cite{latella1towards}.
The counterexamples generated by our PRISM extension DiPro~\cite{AljazzarL09b} 
are subsequently mapped onto fault trees and 
UML sequence diagrams, which makes them interpretable at the level of the UML model. 
Due to the fully automated nature of our approach 
the analysis model and the formal methods used during the analysis are completely hidden from the user.
\begin{figure}[h]
\centering
\includegraphics[width=12cm]{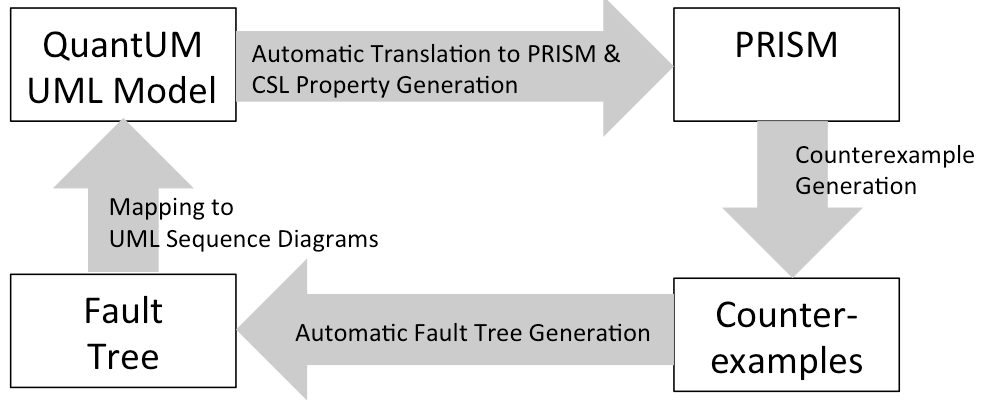}
\caption{Overview of the QuantUM Approach.}
\label{fig:approach}
\end{figure}
Our approach depicted in Fig. \ref{fig:approach} can be summarized by identifying the following steps:
\begin{itemize}
\item Our UML extension is used to annotate the UML model with all
information that is needed to perform a dependability analysis.
\item The annotated UML model is then exported  
in the XML Metadata Interchange (XMI) format \cite{xmi} which is the standard format for exchanging UML models.   
\item Subsequently, our QuantUM Tool parses the generated XMI file and generates the analysis model
in the input language of the probabilistic model checker PRISM as well the properties to be verified.
\item For the analysis we use the probabilistic 
model checker PRISM together with DiPro in order to compute probabilistic counterexamples
representing paths leading to a hazard state.
\item The resulting counterexamples can then be transformed into a fault tree 
that can be interpreted at the level of the UML model. Alternatively, 
they can be mapped onto a UML sequence diagram which can
be displayed in the UML modeling tool containing the original UML model.
\end{itemize}

In order to achieve the full automation that we argued for above it is important 
that a) the properties to be verified are automatically generated, and b) the 
counterexamples can be interpreted at the level of the UML model. We next discuss 
these two issues in more detail.

\section{Automatic Generation of CSL Formulas}\label{sec:CSL}
The properties to be analyzed are important inputs to the analysis of the QuantUM extended UML model. 
In stochastic model checking using PRISM, the property that is to be
verified needs to be specified using a variant of temporal logic called 
Continuous Stochastic Logic (CSL)~\cite{AzizSSB96,BaierHHK03}.
We offer two possibilities for property specification: 
first we automatically generate a set of CSL properties out of the UML model, 
and second we allow the user to manually specify CSL properties. 

We only give a brief introduction into CSL, for a more comprehensive description we refer to \cite{BaierHHK03}. 
CSL is a stochastic variant of the Computation Tree Logic (CTL)~\cite{emerson:86}  
with state and path formulas based on \cite{AzizSSB00}. 
The state formulas are interpreted over states of a continuous-time Markov chain (CTMC) \cite{kulkarni1995modeling}, 
whereas the path formulas are interpreted over paths in a CTMC. 
CSL extends CTL with two probabilistic operators that refer to the steady state and transient behavior of the model. 
The steady-state operator refers to the probability of residing in a particular set of states, specified by a state 
formula, in the long run.
The transient operator allows us to refer to the probability of the occurrence of particular paths in the CTMC. 
In order to express the time span of a certain path, the path operators until ($U$) and next ($X$) 
are extended with a parameter that specifies a time interval.

\begin{figure}[htb]
\centering
\includegraphics[width=8cm]{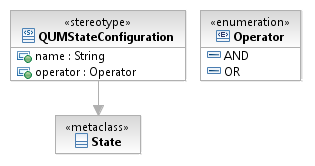}
\caption{Definition of the  \emph{QUMStateConfiguration} stereotype.}
\label{fig:QuantUM_QuantUM_Property}
\end{figure}

The automatic generation of CSL formulas out of the UML model requires the
definition of a UML extension that allows us to refer to states of the UML
model in the generated formulas. 
We introduce the stereotype
\emph{QUMStateConfiguration} (cf. Fig. \ref{fig:QuantUM_QuantUM_Property}) 
as part of our UML Profile described in \cite{mscflf}.
This stereotype can be used to assign names to state configurations. 
In order to do so, the stereotype is assigned to \emph{states} in the state machines
of the UML model.  All \emph{QUMStateConfiguration} stereotypes with the same \emph{name} 
are treated as one state configuration. This is reminiscent of labels in the PRISM language.
A state configuration can also be seen as a boolean formula, each state can either be true when the system 
is in this state or false, when the system is not in this state. The \emph{operator} 
variable indicates whether the boolean variables representing the states are connected by an 
\emph{and}-operator (\emph{AND})  or an \emph{or}-operator (\emph{OR}).
The name of the state configuration is used during model to refer to these state configurations 
in the UML model.

We now discuss how
the states of the UML model are translated into the PRISM model. 
For each \emph{QUMComponent}, its states are encoded as integer values.
States representing the normal behavior are always assigned a value 
between 0 and the total number of states representing normal behavior (\emph{\#normstate}). 
All failure states are identified by a number that is greater than \emph{\#normstate}. 
The placeholders \emph{ \%\#normstate\%} and \emph{ \%\#failstates\%} represent the number of
states in the normal behavior state machine and failure pattern state machines respectively.
The variable  \emph{\%module\_id\%\_state} is then used, in the PRISM model, to represent 
the states of the UML state machines, 
according to the state encoding explained above.
Note that  \emph{\%module\_id\%} will be replaced by an identifier which refers to the respective 
\emph{QUMComponent}. The translation of the model to PRISM is done in linear
time, by iterating over the QUMComponents.

In the following we describe how state formulas can automatically be generated by the QuantUM tool.
The properties that we wish to generate fall into three categories:  
\begin{itemize}
\item
First we generate a property 
that can be used to compute the probability of a component to be in the failed state.
This formula is generated for each component with the stereotype \emph{QUMComponent}.
\item
Second, we generate a formula which computes the probability of the failure of any
component, which means that at least one of the components with the stereotype \emph{QUMComponent} 
is in a failure state. 
\item
Third, we compute one formula for each \emph{QUMStateConfiguration}
that can be used to compute the probability of reaching this state configuration.
\end{itemize}
In the following we describe how formulas falling into these categories can be automatically
generated.

\paragraph{Probability of the failure of a specified \emph{QUMComponent}.}
A \emph{QUMComponent} is failed whenever it has entered a failure pattern state machine. 
Hence, whenever the value of the variable \emph{\%module\_id\%\_state} is greater than 
\emph{ \%\#normstate\%}, the component is failed.
The resulting state formula representing the failure of a component is: 
\[(\%\mbox{module\_id}\%\_\mbox{state} >\%\#\mbox{normstate}\%)\]
Thus, the CSL formula 
\[P_{=?} [ (\mbox{true}) U^{<=T} (\%\mbox{module\_id}\%\_\mbox{state} >\%\#\mbox{normstate}\%)]\]
can be used to determine the probability of a failure of the \emph{QUMComponent} 
with the specified  module id within mission time T.

\paragraph{Probability of the failure of any \emph{QUMComponent}.}
The state formula 
\begin{eqnarray*}
(\%\mbox{module\_id}_{1}\%\_\mbox{state} >\%\#\mbox{normstate}\%) &|& \\ (\%\mbox{module\_id}_{2}\%\_\mbox{state} > \%\#\mbox{normstate}\%) &| ... |& \\  (\%\mbox{module\_id}_{n}\%\_\mbox{state} >\%\#\mbox{normstate}\%)
\end{eqnarray*}
represents the failure of any of the components
with module ids $\%\mbox{module\_id}_{1}\% ... \%\mbox{module\_id}_{n}\%$.
Similarly to the above, the CSL formula 
\begin{eqnarray*}
P_{=?} [ (\mbox{true}) U^{<=T} ((\%\mbox{module\_id}_{1}\%\_\mbox{state} >\%\#\mbox{normstate}\%) &|& \\ (\%\mbox{module\_id}_{2}\%\_\mbox{state} >\%\#\mbox{normstate}\%) &|  ... |&  \\ (\%\mbox{module\_id}_{n}\%\_\mbox{state} >\%\#\mbox{normstate}\%))]
\end{eqnarray*}
can be used to determine the probability of a failure of any of the \emph{QUMComponents}  within mission time T.

\paragraph{Probability of a QUMStateConfiguration.}
As already explained, each \emph{QUMStateConfiguration} can also be interpreted as a boolean formula. 
A state can then either be true, when the module is in this state or in one of its sub-states, 
or false, when the module is not in this state. 
The \emph{operator} variable indicates whether the boolean variables representing the states are connected
 by a boolean \emph{and}-operator (\emph{AND})  or a boolean \emph{or}-operator (\emph{OR}).
The states are identified by the state encoding explained above, hence the boolean expressions
 \[\mbox{in\_state\_id} = (\mbox{state\_id} \le \%\mbox{module\_id}\%\_\mbox{state} \le \mbox{state\_id\_substate}_{n})\] 
can be used to determine whether  
a \emph{QUMComponent} is in a state with the state id in \emph{\%state\_id\%}, or in one of its sub-states. 
If the state with \emph{\%state\_id\%} does not have sub-states, 
the expression \[\mbox{in\_state\_id} = (\%\mbox{module\_id}\%\_\mbox{state} = \%\mbox{state\_id}\%)\] suffices.
To obtain the complete state configuration we connect the individual formulas, identifying the single module states ($\mbox{in\_state\_id}_{1}, ... ,  \mbox{in\_state\_id}_{n}$ ) of the \emph{QUMStateConfiguration},  by the appropriate Boolean operators.
For the \emph{and}-operator we get 
\[\varphi = ( \mbox{in\_state\_id}_{1} \& \mbox{in\_state\_id}_{2} \& ...
\& \mbox{in\_state\_id}_{n}) \] 
as state formula representing the \emph{QUMStateConfiguration},
for the \emph{or}-operator we get \[\varphi= (\mbox{in\_state\_id}_{1} | \mbox{in\_state\_id}_{2} | ...
| \mbox{in\_state\_id}_{n})\] respectively.
In analogy to the previous cases, the CSL formula \[P_{=?} [ (\mbox{true}) U^{<=T} (\varphi)]\]
can be used to determine the probability of reaching the \emph{QUMStateConfiguration}  within mission time T. 

In addition to these three categories, we also generate the state formulas identifying the 
\emph{QUMStateConfiguration}. These can be used to manually build more advanced CSL formulas in
order to express more complicated properties.
We propose to use the ProProST specification patterns~\cite{1368094} in order to obtain 
semantically correct CSL property specifications.
It should be noted that 
according to the ProProST specification pattern system the CSL formulas that we generate all 
fall into the category of 
probabilistic existence. Also note that all formula patterns defined in~\cite{1368094} can be constructed 
with our tool using the state formulas that we synthesize. 

\section{Mapping of Probabilistic Counterexamples onto UML Sequence Diagrams}\label{sec:Sequence}
In order to hide the formal analysis model from the user, 
it is necessary to lift the results of the analysis to the level of the UML model. 
In a first step, we map the counterexamples generated by DiPro 
to Fault Trees \cite{nrc:81}. This mapping is described in detail in \cite{kuntzLeiLeu11}
and is based on counterfactual-type causality reasoning \cite{halpern2005causes}.   
For the purpose of this paper 
it suffices to say
that we establish the causal relationships of the events of the model, that is transitions in the PRISM model,
and the hazard (or other state formula) under consideration. 
This allows us to omit events from the counterexamples that are not causal for the hazard. 
The remaining events are then displayed in a fault tree. Even though the representation of the counterexamples 
as fault tree already lifts the counterexamples onto the level of the UML model and hence 
facilitates the analysis, it is still desirable to display the counterexamples in the same language
as the model that is analyzed, namely UML. We therefore also describe a translation of the counterexample
to UML Sequence Diagrams.

\begin{figure}[ht!]
\centering
\includegraphics[width=5.5cm]{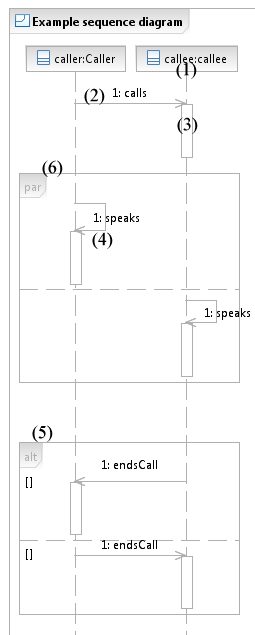}
\caption{Example of a UML sequence diagram.}
\label{fig:sequence}
\end{figure}

Since paths in a counterexample represent sequences of the system executions, UML sequence diagrams
are an obvious choice to display probabilistic counterexamples.
A sequence diagram in the UML is a type of interaction diagram, 
similar to Message Sequence Charts \cite{msc},
that shows how processes interact with one another and in what order.
An example of a UML sequence diagram can be found in Figure \ref{fig:sequence}.

In a sequence diagram different processes or objects that live simultaneously are represented as parallel vertical lines, 
so called lifelines (cf. (1) in Figure \ref{fig:sequence}).
The messages exchanged between those objects are represented by horizontal arrows 
with the message name written above them (cf. (2) in Figure \ref{fig:sequence}). 
Activation boxes, or method-call boxes, are opaque rectangles drawn on top of lifelines to
represent that processes are being performed in response to the reception of a 
message (cf. (3) in Figure \ref{fig:sequence}). 
Objects calling methods on themselves use messages and add new activation boxes on top of any other activation boxes  
to indicate a further level of processing
(cf. (4) in Figure \ref{fig:sequence}). 
The order of events along individual lifelines is total, from top to bottom.
The order of potentially concurrent events across different lifelines is partial, in keeping with
the happened-before relation of~\cite{Lam78}.
In order to impose a total order of events, message names can be labeled with natural numbers, as 
illustrated in Figure \ref{fig:AirbagUML1}. 
Combined interaction fragments can be used to change this event order and to 
display alternative (\emph{alt}) sequences (cf. (5) in Figure \ref{fig:sequence}) or sequences that run concurrently (\emph{par}) 
(cf. (6) in Figure \ref{fig:sequence}). 

Theoretically it is possible to map all paths of the counterexample directly to the sequence 
diagram. Since this mapping would result in a sequence diagram with hundreds of
alternative sequences, it is necessary to extract those events from the counterexample 
that are causal for the hazard. This extraction is done by our fault tree computation algorithm.
The paths of the counterexample returned by this algorithm are alternative execution paths of the system with the probability P. 
In order to express alternative executions in sequence diagrams, an \emph{alt} combined interaction fragment is used. 
The probability of the path is given in the name of the corresponding \emph{alt} combined interaction fragment.
For each \emph{QUMComponent} of the model we add a lifeline representing the component.
We add a function \emph{transition("source state","target state")} to each component, in order to visualize the
transitions that are taken inside the component. 
A call to the \emph{transition} function is added for each transition in the fault tree. 
If the transition is a call of an operation of a \emph{QUMComponent},
an operation call from the caller component to the callee component is added to the sequence diagram.
If the paths of the counterexample would be mapped directly to the sequence diagram, all paths, 
including possible interleavings of the paths, would be added to the sequence diagram.  
When the fault tree computation is used to filter the paths, it is checked whether the order of the events does have effect on the causality.
If the order of the events does not have an effect on the causality, the events are represented by \emph{par} combined fragments 
where each compartment represents one event. The compartments 
of the \emph{par} combined fragment can be either executed in parallel or in any other possible concurrent interleaving. 
Otherwise, that is the order of the event is relevant, the events are mapped to a sequence representing their order. 

In order to display the sequence diagrams, the QuantUM tool appends the XMI code of the diagram to the XMI file that contains
the UML model. Thus, it can easily be imported in the UML CASE-tool that was used to edit the original UML design model. 

\section{Case Study}\label{sec:CaseStudy}

We have applied our modeling and analysis approach to a case study from
the automotive embedded software domain. We performed an analysis of the design
of an Electronic Control Unit for an Airbag system that is being developed 
at TRW Automotive GmbH, see also \cite{AljazzarL09a}. 
Note that the used probability values are merely approximate "ballpark" numbers, since the actual  
values are intellectual property of our industrial partner TRW Automotive GmbH 
that we are not allowed to publish.

The airbag system architecture that we consider consists of
two acceleration sensors whose task is to detect front or rear crashes,
one microcontroller to perform the crash evaluation,
and an actuator that controls the deployment of the airbag.

The deployment of the airbag is secured by two redundant protection mechanisms.
The Field Effect Transistor (FET) controls the power supply for the airbag 
squibs. If the Field Effect Transistor is not armed, which means that
the FET-Pin is not high,
the airbag squib does not have enough electrical power to ignite the airbag.
The second protection mechanism
is the Firing Application Specific Integrated Circuit (FASIC) which controls the airbag squib.
Only if it receives first an arm command and then a fire command from the microcontroller
it will ignite the airbag squib.

\begin{figure}[htb]
\centering
\includegraphics[width=\textwidth]{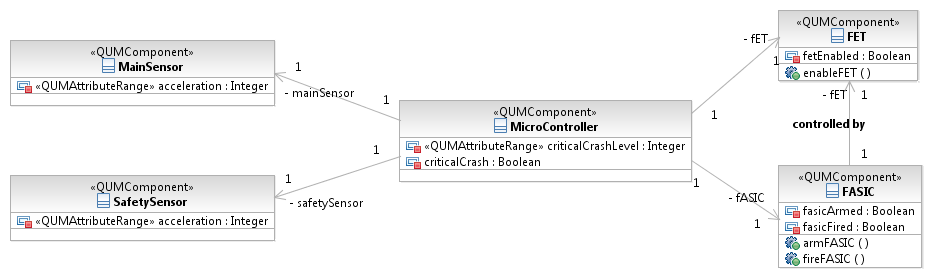}
\caption{Class diagram modeling the airbag system.}
\label{fig:AirbagSys}
\end{figure}

Although airbags save lives in crash situations, they may be fatal 
if they are inadvertently deployed.
This is because the driver may lose control of
the car when this accidental deployment occurs.
It is hence a pivotal safety requirement
that an airbag is never deployed if there is no crash situation.
In order to analyze how safe the considered system architecture,
modeled with the CASE tool IBM Rational Software Architect and shown in Figure \ref{fig:AirbagSys},
is, we annotated the model with our QuantUM extension and 
performed an analysis with the QuantUM tool. 

\begin{figure}[htb]
\centering
\includegraphics[width=\textwidth]{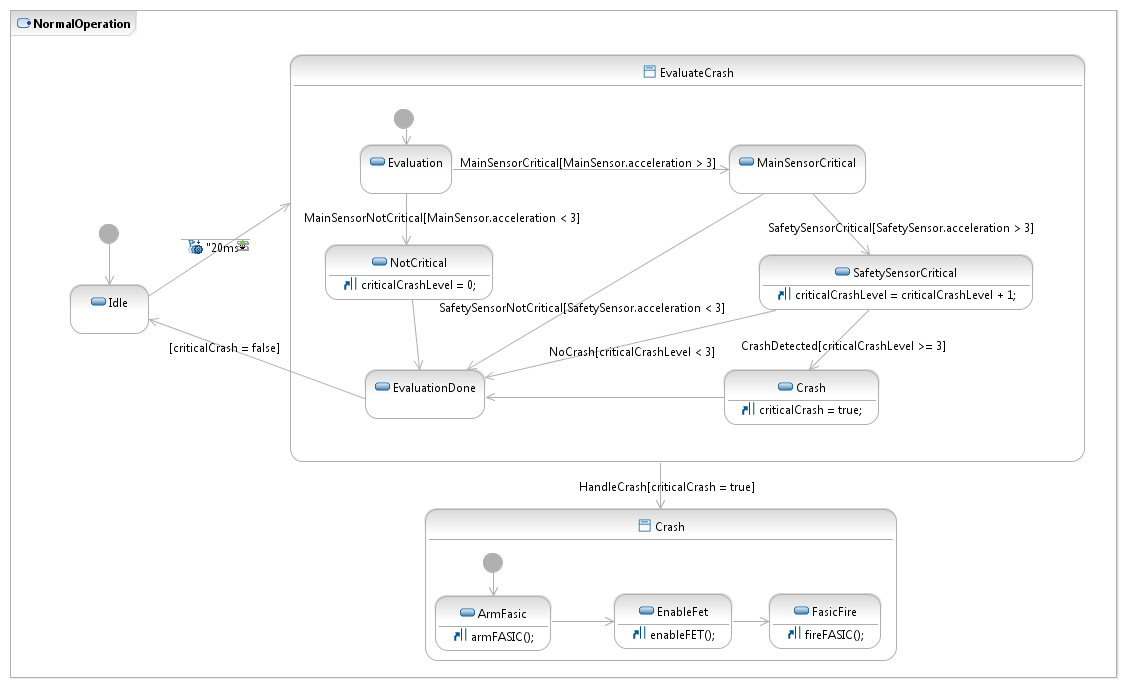}
\caption{State machine representing the normal behavior of the microcontroller.}
\label{fig:AirbagSystem_NormalOperation}
\end{figure}

\begin{figure}[htb]
\centering
\includegraphics[width=10cm]{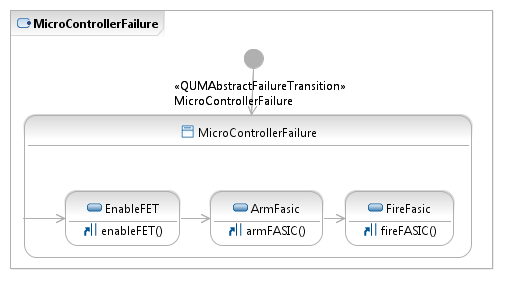}
\caption{State machine representing the failure pattern  of the microcontroller.}
\label{fig:AirbagSystem_MicroControllerFailure}
\end{figure}

The \emph{MicroController} \emph{QUMComponent} for instance, comprises one state machine representing its
normal behavior (cf. Fig. \ref{fig:AirbagSystem_NormalOperation}) and one
state machine representing the failure pattern, which is shown in Fig. \ref{fig:AirbagSystem_MicroControllerFailure}.
If the \emph{MicroController} is in the normal state machine, 
it evaluates every 20ms whether there is a crash or not.
If the \emph{MainSensor} and the \emph{SafetySensor} were giving readings above the threshold ($\mbox{ \em acceleration} > 3$) 
for three consecutive evaluations, the \emph{MicroController} concludes that there is a crash and the state $Crash$ is entered. 
In the \emph{Crash} state, the \emph{FASIC} is first armed, then the \emph{FET} is enabled and  
finally, the \emph{FASIC} is fired. The failure pattern of the \emph{MicroController} can be entered at any time, 
with the rate specified for the \emph{QUMAbstractFailureTransition}. 
As soon as it is entered it will execute the fire sequence: 
enable \emph{FET}, arm \emph{FASIC}, and fire \emph{FASIC}. 

After all annotations were made, we exported the model into an XMI file,
which was then imported by the QuantUM tool and translated into a PRISM model.
The import of the XMI file and translation of the model was completed in less than two seconds. 
Without the QuantUM tool, this process would require days of work of a trained engineer.

The resulting PRISM model consists of 3249 states and 15390 transitions. 
The QuantUM tool also generated the CSL formula 
$P_{=?} [(\mbox{true}) U^{<=T} (\mbox{\em inadvertent\_deployment})]$
where \emph{inadvertent\_deployment} is replaced by the state formula 
which identifies all states in the \emph{QUMStateConfiguration} with the value \emph{inadvertent\_deployment}. 
\emph{T} represents the mission time. 
Since the acceleration value of the sensor state machines is always zero, 
the formula $P_{=?} [(\mbox{true}) U^{<=T} (\mbox{\em inadvertent\_deployment})]$ calculates 
the probability of the airbag being deployed, during the mission time $T$, 
when there is no crash situation.
Owing to the use of the QuantUM tool, the only input which has to be given by the user is the mission time $T$.

We computed probabilities for the mission times T=10, T=100, and T=1000
and recorded the runtime for the counterexample computation (Runtime CX),
the number of paths in the counterexample (Paths in CX), the runtime of
the fault tree generation algorithm (Runtime FT) and the numbers of paths in the fault tree
(Paths in FT) in Figure \ref{fig:TableAirbagModel}. The experiments where performed 
on a PC with an Intel QuadCore i5 processor with 2.67 Ghz and 8 GBs of RAM. 

\begin{figure}[htb]
\footnotesize
\centering
\begin{tabular}{lllll}
\toprule
 T & Runtime CX (sec.) & Paths in CX  & Runtime FT (sec.) & Paths in FT\\
\midrule
 10 &  646.425 (approx.  10.77 min.) & 738 & 2.86 & 5 \\
\midrule
100 & 664.893 (approx. 11.08 min.)&  738 & 3.52 & 5\\
\midrule
 1000& 820.431 (approx. 15.67 min.)& 738 & 2.98 & 5\\
\bottomrule
\end{tabular}
\caption{Experiment results for T=10, T=100 and T=1000.}
\label{fig:TableAirbagModel}
\end{figure}

Figure \ref{fig:TableAirbagModel} shows that the computation of the
fault tree is finished in several seconds, whereas the computation of
the counterexample takes several minutes. While the different 
running times of the counterexample computation algorithm
seem to be caused by the different values of the mission time $T$,
the variation of the running time of the fault tree computation seems to be caused by
background processes on the PC on which the experiments were conducted.

\begin{figure}[htb]
\centering
\includegraphics[width=\textwidth]{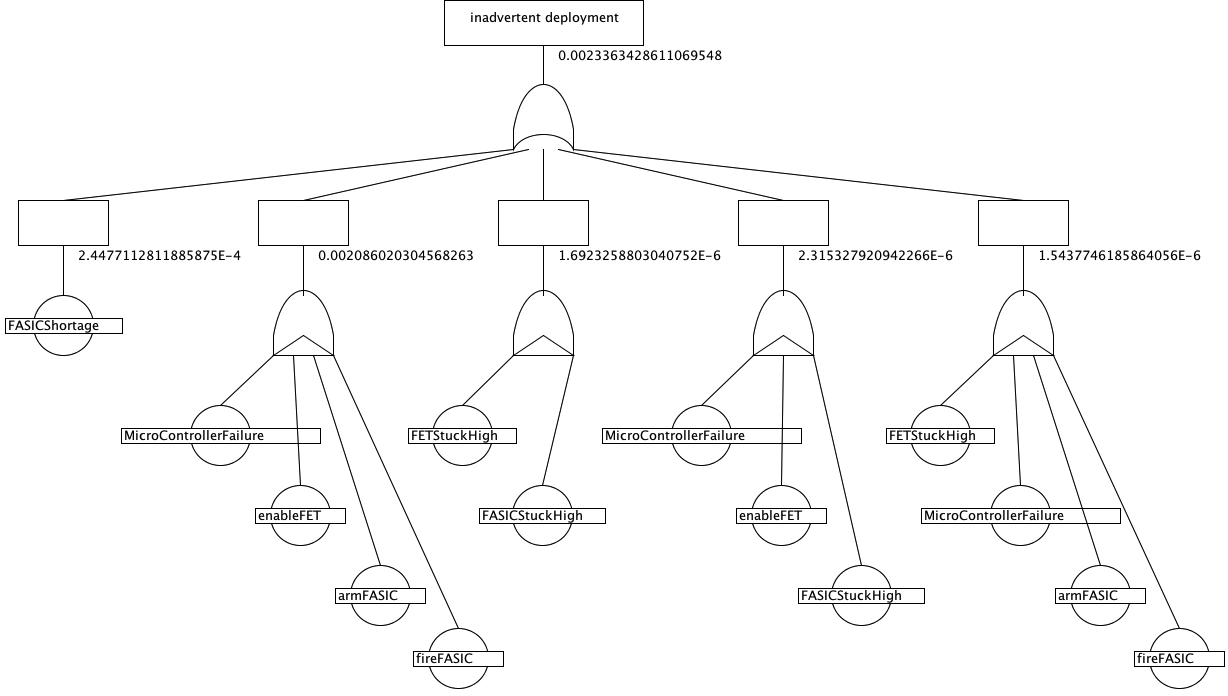}
\caption{Fault tree for the \emph{QUMStateConfiguration} \emph{inadvertent\_deployment} (T = 10).}
\label{fig:AirbagFT}
\end{figure}

Figure \ref{fig:AirbagFT} shows the fault tree generated from the counterexample 
for T=10. 
While the counterexample consists of 738 paths, the fault tree comprises only 5 paths.
It is easy to see by which basic events, and with which probabilities, 
an inadvertent deployment of the airbag is caused. 
Obviously, there is only one single fault that can lead to an inadvertent deployment, namely
\emph{FASICShortage}. It is also easy to recognize that the basic event 
\emph{MicroControllerFailure}, for instance, can only lead to an inadvertent deployment 
if it is followed by one of the following sequences of basic events: \emph{enableFET},
\emph{armFASIC}, and \emph{fireFASIC} or \emph{enableFET}, and \emph{FASICStuckHigh}.
If the basic event \emph{FETStuckHigh} occurs prior to the \emph{MicroControllerFailure}, then 
the sequence \emph{armFASIC} and \emph{fireFASIC} occurring after the \emph{MicroControllerFailure}
event suffices.

The case study shows that the fault tree is a compact and concise visualization
of the counterexample which allows for an easy identification of the basic events
that cause the inadvertent deployment of the airbag, and their corresponding probabilities.
If the order of the events is important, this can be seen in the fault tree by the \emph{PAND}-gate.
Using a manual analysis 
one would have to manually compare the order of the events in all 738 paths in the counterexample,
which is a tedious and time consuming task.  

\begin{figure}[htb]
\centering
\includegraphics[width=\textwidth]{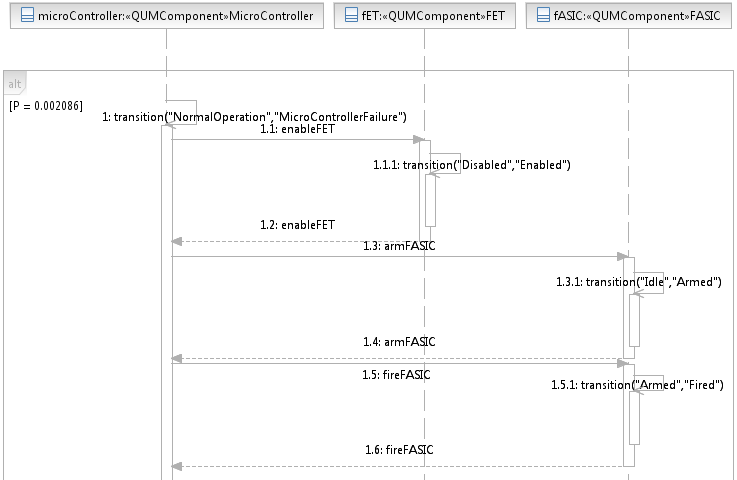}
\caption{Fragment of the UML sequence diagram for the \emph{QUMStateConfiguration} \emph{inadvertent\_deployment} (T = 10).}
\label{fig:AirbagUML1}
\end{figure}

Figure \ref{fig:AirbagUML1} shows a fragment of the UML sequence diagram
which visualizes the counterexample for T=10.
The generation of the XMI code of the sequence diagram took less then one second. 
We imported the XMI code into the UML model of the airbag system in the CASE tool IBM Rational Software Architect.
This allows us to interpret the counterexample directly in the CASE tool. 
An additional benefit of the visualization of the counterexample as a sequence diagram 
is that operation calls can be depicted. In Figure \ref{fig:AirbagUML1}, for instance, it is easy to see how after a failure of the microcontroller
the operations \emph{enableFET()}, \emph{armFASIC()}, and \emph{fireFASIC()} are called.

\section{Related Work}\label{sec:RelatedWork}
The idea of using UML models to derive models for quantitative safety analysis is not new.
In \cite{majzik2003stochastic} the authors present a UML profile for annotating software
dependability properties. This annotated model is then transformed into an intermediate model,
that is then transformed into Timed Petri Nets. The main drawbacks of this work is that it merely
focuses on the structural aspect of the UML model, while the actual behavioral description is not considered. 
Another drawback is the introduction of unnecessary redundant information in the UML model, 
since sometimes the joint use of more than one stereotype is required.
In \cite{bernardi-dependability} the authors extend the UML Profile for Modeling and Analysis of Real Time Embedded Systems 
(MARTE)~\cite{marte} with a profile for dependability analysis and modeling. While this work is very expressive,
it heavily relies on the use of the MARTE profile, which is only supported by very few UML CASE tools.
Additionally, the amount of stereotypes, tagged values and annotations that need to be added to the model
is very large.  
Another drawback of this approach is that the translation from the annotated UML model into the 
Deterministic and Stochastic Petri Nets (DSPN) \cite{marsan1987petri} used for analysis is carried out 
manually which is, as we argue above, an error-prone and risky task for large UML models.
The work defined in \cite{jansenPHD03} presents a stochastic extension of 
the Statechart notation, called StoCharts. 
The StoChart approach suffers from the following drawbacks. First, it is restricted to the analysis
of the behavioral aspects of the system and does not allow for structural analysis. Second,  
while there exist some tools that allow to draw StoCharts, there is no integration of StoCharts
into UML models available.
In \cite{boudali:08b} the architecture dependability analysis framework Arcades is presented. 
While Arcade is very expressive and was applied to hardware, software and infrastructure systems,
the main drawback is that it is based on a textual description of the system and hence would require a manual 
translation process of the UML model to Arcade.

We are, to the best of our knowledge,  
not aware of any alternative approach that allows for the automatic generation of the analysis model and the
automatic CSL property construction.

\section{Future Work}\label{sec:FutureWork}
In future work we plan to extend the expressiveness of the QuantUM profile, 
to integrate methods to further facilitate automatic stochastic property specification, 
and to apply our approach on other architecture description languages such as, for instance, SysML \cite{sysmlspec}.

At the moment the approach can be applied to all system and software architecture models  
that do not make use of complex programming code in the entry-, during- or exit-action of the states.
To further extend our approach we plan to work on methods for the probabilistic 
analysis of object-oriented source code attached to state machine transitions. 

Additionally, we plan to conduct more case studies to prove the scalability of our 
approach.

\section{Conclusion}\label{sec:Conclusion}
We have presented an UML profile that allows for the annotation of UML models with quantitative information,
together with a tool that automatically translates the model into the PRISM language and performs the analysis with PRISM. 
In addition to the translation of the model, the tool also allows for the automatic generation of CSL properties.
Furthermore, we have developed  a method, not described in detail in this paper, 
that automatically generates a fault tree from a 
probabilistic counterexample. The resulting fault trees were significantly smaller than the 
probabilistic counterexamples and hence easier to understand, while still presenting 
all causally relevant information.
We also presented a mapping of probabilistic counterexamples
to UML sequence diagrams which make the counterexamples interpretable inside
the UML modeling tool. We believe that this is particularly useful for  system 
and software architects which usually can interpret an UML sequence diagram 
much better than fault trees, whereas safety engineers might prefer the visualization
of the counterexamples as fault trees.
We illustrated the usefulness of our modeling and analysis approach using an industrial case 
study.

\nocite{*}
\bibliographystyle{eptcs}
\bibliography{literature}
\end{document}